\documentclass[journal]{IEEEtran}

\usepackage{cite}
\usepackage{hyperref}
\hypersetup{
    colorlinks=true,
    linkcolor=blue,
    filecolor=magenta,      
    urlcolor=cyan,
    citecolor=blue,
}
\usepackage[table]{xcolor}

\ifCLASSINFOpdf
  \usepackage[pdftex]{graphicx}
  \graphicspath{{./img/}}
  \DeclareGraphicsExtensions{.pdf,.jpeg,.png}
\else
\fi

\usepackage[utf8]{inputenc}

\begin{document}
\title{Search techniques in peer to peer networks}

\author{Mohamed~Elsharnouby,~\textit{mohamedelsharnouby@std.sehir.edu.tr}\\
        \textit{Istanbul Sehir University}
}

\maketitle

\begin{abstract}
Peer to peer (P2P) networks are an overlay on IP network of the internet and they can shape the future of computing by their involvement in distributed systems with the increased of use of low priced personal computers to form big clusters of distributed systems. An important problem for P2P networks and models is searching for data in the network which can be the basis of any service that uses such a network. Here we explore the major types of P2P networks and their solution to such a problem and we explore improvements that happened to these networks to be more appealing for commercial usage by offering features of load balancing, scalability, self organization and fault tolerance.
\end{abstract}

\begin{IEEEkeywords}
p2p, peer-to-peer, search, networks.
\end{IEEEkeywords}

\section{Introduction}
\label{sec:intro}

\IEEEPARstart{P}{eer} to peer networks are not just used in popular file sharing networks but they are also the basis for many distributed systems as in distributed storage, caching and load balancing systems. Peer to peer (P2P) networks are just considered an overlay network as peers self organize on top of the Internet Protocol (IP) networks to offer a specific functionality as routing, search for data, knowledge about nearby peers, anonymity, fault tolerance and redundant storage. The idea of peer to peer networks became so popular after the famous file sharing network Napster which was a music sharing network that allowed users to share music in between each other. But Napster had a critical weakness that contributed to its failure which was that it depended on a central server on which there was an index of all the files that users are sharing and when a user requires a file he contacts that server to get the IP of the user who has that file and then the download starts from that provider. The idea of having a central server with indices of all shared files made a single point of failure which contributed to the fall of the network after shutting down that server. After that multiple peer to peer networks appeared as Gnutella, BitTorrent and others which we will discuss later. Each of those late networks had their own weakness and strong points which made some of them better than others for certain services.

\section{Searching for data in P2P networks}
Searching for data in peer to peer networks depends on having a Distributed Hash Table (DHT) which is considered the foundation of most searching algorithms. DHT depends on that each data item is assigned a key and the value for that key is the peer who has that data item and the task of any algorithm is to implement the main operation of any DHT which is \textit{lookup(key)} which takes a data item key as an argument and returns the value as the location (IP address for example) of the providing peer who has that data item so that the requester can download it from the provider. The same operation could be used as well for inserting data items in a peer to peer network, this is done by generating a hash key for the item that the publisher peer wants to store (a hash function as SHA-1 could be used for that) and calls the \textit{lookup(key)} function using the generated key which will yield one of the peers who is willing to store that data item. Such a distributed storage system should handle other operations as replication, caching and other issues. The mentioned scenario is just for a distributed data storage system but peer to peer networks could be used for many other services to get benefits from the capabilities they provide.

\section{Peer to peer networks types}
Peer to peer networks could be split into two groups according to their structure which are: Structured networks and Unstructured networks \cite{Balakrishnan2003Looking,Lua2005Survey}. Structured networks are the ones that have a certain systematic or hierarchical structure to make it easier to route requests for data inside the network and find data more efficiently compared to unstructured P2P networks. Examples of these networks are Content Addressable Network (CAN) \cite{Ratnasamy2001Scalable}, Chord \cite{Stoica2001Chord}, Tapestry \cite{Zhao2004Tapestry}, Pastry \cite{Rowstron2001Pastry}, Viceroy \cite{Malkhi2002Viceroy} and others.

Unstructured networks are the ones that don't depend on a certain structure of the network for searching for data inside the network to work properly, this might give the network a better advantage than structured in terms of being more resilient and better anonymity of peers. Examples of those networks are Freenet \cite{Clarke2001Freenet}, Gnutella and BitTorrent.

\subsection{Structured networks}

\subsubsection{Content Addressable Network (CAN)}

\begin{figure}[ht]
\centering
\includegraphics[width=0.4\textwidth]{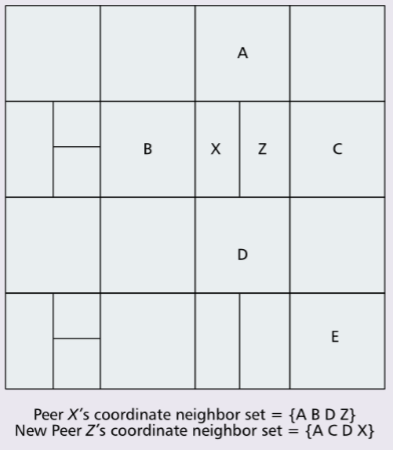}
\caption{Showing CAN with splitted zones in between peers}
\end{figure}

Content Addressable Network (CAN) \cite{Ratnasamy2001Scalable} is structured as a multidimensional Cartesian coordinate space on a multi-torus which virtually maps every peer to a certain zone on the d-dimensional space. In any point of time the whole space is partitioned and distributed as zones on the clients that are present on the network. Every peer is responsible for a certain zone if that space and stores a table containing the IP addresses and the coordinates of the zones of the surrounding peers in that space.

Routing in this network happens for a message that contains the coordinates of the target zone by using the shortest path in this Cartesian coordinate space. Any peer receiving a message would check greedily for the closes peer out of its neighbors that gets closer to the target zone and forwards the message to that neighbor. The routing performance of this algorithm simply depends on the number of dimensions of the space and has performance of $O(d \times N^{1/d})$ where d is the number of dimensions and N is the number of peers in the network.

A new peer joining the network will only need to be assigned a zone and this is done using a uniform random function to choose a point in the space then a message is routed through the network till it reaches the peer that is responsible for this zone that contains the chosen point. That peer then divides the zone that it was responsible for into half and a message updating the routing tables of the neighboring peers is sent, it also learns about the surrounding peers from the old peer. Also the key and value pairs from half of that zone gets handed over to the new joining peer.

A leaving peer leaves an empty zone which is known when a periodic message that's being sent by each peer to its neighbors is absent for sometime. When this happens a takeover algorithm ensures that the empty zone either gets merged with a neighboring zone to form a valid zone or just joined with the smallest zone of the surrounding neighboring peers. Then the neighboring peers are updated about that change to update their routing tables.

Many improvements could be integrated into CAN to increase the routing performance and increase the robustness of the network against multiple peers failures. First, the dimensions of the space could be increased and that will greatly reduce the routing distance but would be to a constant factor. The increase of dimensions will increase robustness as well as that increased the number of neighboring peers for each peer which increases the number of alternate paths that could be taken by a request in case if peers' failures. Second, a multiple coordinate spaces (Realities) concept could be used in which there are many coordinate spaces and every peer is assigned a different zone in every coordinate space (reality). Also, every data item is put in the same point in all realities which increases the replication and availability of data in realities which means that only this data item will be lost if all peers in all realities on that point in the space fail together. Routing also is improved using this as a peer can route using the reality in which it is closest to the target point. The original paper compared the two techniques of increasing dimensions and increasing realities and which of those provided better improvement for CAN and they used round-trip-time metric (RTT) for better assessment of those improvements. It was found out that increasing the dimensions of space gives better improvements in terms of path shortness but multiple realities still has the advantages of improved data availability and fault tolerance.

A third improvement was overloading coordinate zones in which every zone is being responsible from multiple peers which reduced path length as it decreased the total number of zones in the system, reduced per-hop latency as a peer now has multiple options as neighbors which could be chosen according to the RTT metric and improved fault tolerance as a zone will be empty only if all peers in the same zone crashed. Fourth, using multiple hash functions could assign every data item multiple places on the coordinate space which will improve data availability. Other improvements included topological awareness of the underlying IP network while constructing CAN but it resulted in a non uniform population of the network, caching and replication of data items to improve the performance of the network. It seems that using a bare bones CAN network results in a very poor performance due to its almost linear routing algorithm. But using a CAN along with all mentioned improvements can results in a network that achieves comparable results in terms of performance to Chord, Tapestry and Viceroy considering the scale of real life applications plus the additional fault tolerance benefits that resulted from those improvements.

\subsubsection{Chord}

Chord \cite{Stoica2001Chord} is structured as an identifier circle on which peers are placed. It depends on consistent hashing to assign each peer a certain ID which lies on the identifier circle and also data items are assigned same IDs on the circle. Every peer has a routing table (Finger table) which is carrying the address of nodes that lie on the circle but are within a constant to successive powers as for a node with ID identifier, it carries those IDs in its routing table: $\{(ID + 2^0), (ID + 2^1), (ID + 2^2), (ID + 2^3), (ID + 2^4), ...\}$. Also, every node carries a number $r$ of successive nodes' IDs in case that a the node cannot route a message through its Finger table. Every data item with IDs on the circle are assigned to the successor node on the circle. A successor is the first node that lies directly at or after the location of that ID of the data item.

Routing in the network happens through sending a message containing the ID of the requested data item and every node picks the longest hop on its Finger table that its ID is still less than the target in the message. This assures the property that a message doesn't exceed or bypass its target while being routed. Having those long connections in the finger table makes the performance of routing in the graph $O(\log N)$.

A joining node gets assigned a new ID on the circle then it gets to handle all data items that are preceding to its ID and the successors in Finger tables of some peers need to be updated to ensure the correctness of routing through the network. This update to Finger tables is ensured by a stabilization protocol which runs in the background periodically to ensure the correctness of Finger tables. Also having the successor keys of $r$ peers that we mentioned before should assure the correctness of routing if any routing is needed after a new peer joined and before Finger tables get updated. Same things happen when a node leaves the network, the stabilization algorithm corrects Finger tables and the data items are handled to the next peer to the leaving one.

\begin{figure}[ht]
\centering
\includegraphics[width=0.4\textwidth]{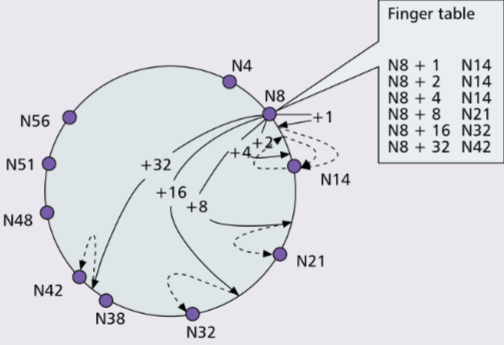}
\caption{Chord Finger table and routing a request using it to find the successor of the data item}
\end{figure}

The main contribution of Chord compared to other type of P2P networks is that it provides a simplified model with very good performance considering routing and it has provable correctness and performance even in so heavy load of peers joining and leaving the network. The downsides of Chord is that it still doesn't have a mechanism to connect partitioned rings back which might happen with many interruptions in the internet connection of many nodes. Also a set of malicious peers could still pose a threat to partitioning a Chord network. Chord also doesn't have a way to optimize for low latency peers to route messages to instead of high latency ones as there was in CAN. Chord doesn't have the same amount of features that could be added to it to increase its robustness as there was in CAN as well. We think that Chord could have the same concept of realities in CAN to increase the routing speed and add more robustness to the network. If every peer is present in a different circles with different IDs (this would require different hashing functions though) this would increase robustness and it will certainly improve routing performance as well as a peer can choose the circle that it's closer to the data object to start routing in.

\subsubsection{Tapestry}
Tapestry \cite{Zhao2004Tapestry} is a lot like Pastry but it takes into account network locality and replication of data items for better routing performance and data availability. Tapestry uses a variation of the famous routing algorithm of Plaxton \cite{Plaxton1997Accessing} which was the basis of many efficient routing algorithms of P2P networks but it didn't take into account the dynamics of a P2P network as it only considered a static network without joining and leaving peers. The concept of the routing algorithm is that every peer is assigned an ID in base $b$ and data is assigned keys in the same hashing space. The routing then happens by every peer by prefix matching. Every peer carries a routing table for each digit in its ID a set of peers that their IDs' prefix match till that digit and they differ in that digit in a way that the peer has all possible digits in that position in its routing table. When a message reaches a peer with a target ID then that peer forwards it to the peer that increases the length of matching prefix with target of the message. The routing performance for this algorithm is $O(\log_b N)$.

An important feature in Tapestry is that it can provide multiple roots to every data item so that it becomes fault tolerant and a lot more robust. Compared to CAN, Tapestry has a much faster routing algorithm also CAN doesn't have the same locality measures and introspective mechanisms that are provided by Tapestry to increase the routing efficiency a lot. Chord cannot make use of real network distances as well in its structure for faster routing. The main advantage of CAN over Tapestry is the simplicity of its joining peer procedure that it can handle a more dynamic network with many joining and leaving nodes easily compared to Tapestry.

\begin{figure}[ht]
\centering
\includegraphics[width=0.4\textwidth]{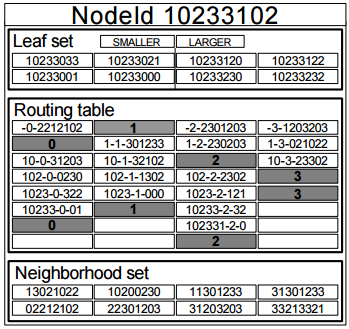}
\caption{Pastry and Tapestry have the same routing tables in which every peer has IDs that differ in a certain digit place and routing happens by increasing the length of the matching suffix with the target ID}
\end{figure}

\subsubsection{Pastry}

Pastry \cite{Rowstron2001Pastry} has the same routing algorithm as Tapestry but it doesn't take into account the locality measures of peers and hence it doesn't optimize for better latency routing paths as Tapestry does. Also, for data replication Pastry replicates data items without control of the user. The last thing is that Pastry replicates the data item around the node which is closest to the data item ID while Tapestry replicates the data item in hops on the path from the requesting peer to the providing peer.

\subsubsection{Viceroy}
Viceroy \cite{Malkhi2002Viceroy} is one of the P2P overlay networks that approximates butterfly networks. Viceroy has a set of connections for each node:
\begin{enumerate}
  \item General Ring: every node is connected to its successor and predecessor
  \item Level Ring: every node is connected to others on the same level in a ring
  \item Butterfly: for every level $L$ for a node there's a "down right" edge that is added to a long range contact in the next level $(L+1)$ at a distance that's around $l/2^l$ away from the node and a "down left" edge which is added to a short range contact on the next level as well $(L+1)$. Also, there's an "up" edge which is added to a close node from the level $(L-1)$ if $L$ is not the first level.
\end{enumerate}
Routing happens by following first the up links in the network then it starts following down links taking left or right according to if the target ID exceeds $l/2^l$ or not until it reaches a level with no more down links which makes the message in the vicinity of the target peer, so the message reaches its target using General ring and Level ring links. The routing algorithm performance of Viceroy is $O(\log N)$.

\subsection{Unstructured networks}

\subsubsection{Freenet}
Freenet \cite{Clarke2001Freenet} is a dynamic P2P network that adapts to the dynamics of the network. It puts anonymity as a prominent property as well compared to other networks. Every peer keeps a dynamic routing table that has the addresses of other peers and the data items that are kept at those peers. The request for a data item is passed on from peer to peer using a Steepest Ascent Hill Climbing with backtracking algorithm in which every peer decides based on the local information it has the next peer that the message should be forwarded to to get closer to the target. Backtracking allows the request to take different paths until the data item is found. A maximum hops limit (Hops to live - HTL) could be assigned for the network to prevent requests from wandering around the network forever. When the data item is found it is cached along the path of the from the requester to the provider for faster access in the future.

The basic caching replacement algorithm used in Freenet is LRU (Least Recently Used) which discards the least recently used item if the cache gets filled. It is used as well for replacement of peers in the routing table when it gets full.

A new cache replacement mechanism was suggested \cite{Zhang2002Using,2001PeertoPeer} to improve the performance of Freenet network. This is done by using the small world concept to develop a cache that has the same properties of a small network as keeping high clustering coefficient while having short paths with few number of hops. That was done by choosing the farthest node $v$ in the routing table in terms of distance of the key space and then comparing the distance of the node $u$ that is to be added to the routing table but it is full. If that distance of u is less than or equal to distance to v then an entry of $u$ is added to cache and $v$ is removed, this ensures high clustering in the network. If the distance of $u$ is more than $v$ then a cache entry of $u$ is to replace $v$ with a certain probability $p$. This concept when compared to the basic LRU cache replacement mechanism it yielded much better performance and resilience of the network to nodes failures. An open question still remains for what is the best value of $p$ to achieve the optimum performance of this cache replacement mechanism for Freenet networks.

\subsubsection{Gnutella}
Gnutella uses a simple Breadth First Search (BFS) for routing requests through the network with a limit on the number of hops for the request. This method is sometimes called flooding as it floods the network with requests until the data item is found. This search mechanism is sadly not scalable and it is said that after the shutdown of Napster the Gnutella network collapsed under its own load because of the huge number of users who joined the network and flooded it with requests.

As every peer in Gnutella is considered a client and a server at the same time and because of the randomness in the construction of the network, it is a very resilient network and is highly fault tolerant. A few improvements have been added to the network later on as the concept of super peers which are peers with better bandwidth connection and they mostly function as central points for routing requests between other peers, this increased the efficiency of the network. Also, if a new peer joins the network it can select some of those ultra peers and publish its own files to them so that they can be shared with other peers more efficiently.

\subsubsection{BitTorrent}
BitTorrent could be considered a centralized unstructured network as it doesn't follow a certain hierarchy in whole but it depends on Trackers which are mostly central servers but even if a few trackers fall down peers can discover each other using the DHT routing tables that are cached in peers' clients. The download of a file in the network is initiated by contacting the set of trackers that are responsible for this torrent file to get the set of peers that are uploading (seeding) the requested file. Then the torrent client connects to those other peers directly to start downloading the file. Any file shared on BitTorrent is cut into small fixed size pieces (256 Kbytes each) to make it easier to track and validate the downloaded file while it is still being downloaded. Whenever one of those pieces of the file is downloaded this peer can now start seeding it to other clients as well. This way the load is distributed on the peers in the network and rare pieces could be downloaded more efficiently as popular ones are distributed over underloaded peers.

BitTorrent implements some algorithms that could prevent free-riders (peers which download only and doesn't upload) by throttling download speeds of those peers and increasing the capacity of peers which contribute more. The robustness of the network of BitTorrent comes from the easiness of starting new trackers which could handle sharing any old files that were shared on other trackers that went down. And the use of DHT for peers to discover other seeds without the need of a tracker using the tables that are cached on some of the already connected peers.

\section{Discussion}
There are still many open questions considering P2P networks \cite{Ratnasamy2002Routing}. Instead of comparing P2P networks to each other we can try asking the questions of can we build a P2P network that combines the best of all networks? An example is in CAN routing is linear while in most of other networks routing is logarithmic, on the contrary, updating neighbors for a joining or leaving peer in CAN is $O(1)$ while in other networks it is mostly $O(\log N)$. So, can we achieve the logarithmic routing along with a constant performance for joining and leaving nodes?

Another question is routing hot spots which means that some peers in the network has so much routing requests compared to other peers. This is different from a peer that has a high load of requests which means it is a target of many requests which can be solved easily. But solving load on intermediate peers which handles so much traffic is a harder problem as there is no way still to reroute certain requests away from those overloaded peers.

In the original paper, there are so many open questions, some of them were solved in some of the P2P networks as incorporating geo locations into consideration while constructing a P2P network which is done by Tapestry although it has its fallbacks. Or making use of the heterogeneity of peers which was solved in Gnutella by the concept of ultra peers to make use of more powerful peers to give the network more efficiency in handling requests. Those techniques are used still in individual P2P networks but those methodologies were not incorporated yet into other networks. For instance Chord can make use of the concept of ultra peers as peers that are distributed on a certain interval around the ID circle as a way of increasing the resilience of the network and provide faster recovery from failing nodes and make the routing a lot faster if most of nodes in the network have an entry for all ultra peers in the network in their Finger table. Also the mentioned before multiple realities that was suggested in CAN but could be implemented in other P2P networks as well.

\section{Concluding remarks}
This paper discussed the majority of P2P networks including structured and unstructured networks and compared their performance in consideration to each other and explored lots of improvements that were done for those networks to increase their performance and make them more appealing for real life commercial use. These P2P networks now make the base for many applications as distributed file storage systems, content delivery networks (as PeerCDN), distributing tasks for a distributed processing system and many others. Finally we had some suggestions and open questions that could be answered in future work or by other researchers.



\bibliography{citeulike}
\bibliographystyle{IEEEtran}

\end{document}